\documentclass{article}
\usepackage[utf8]{inputenc}
\usepackage[T1]{fontenc}
\usepackage{ucs}
\usepackage{amsthm,amsmath,amsfonts,amssymb}
\usepackage{pdfsync}
\usepackage{natbib}
\usepackage{enumerate}
\usepackage{graphicx}
\usepackage{epsfig} 
\usepackage{calc}
\usepackage{endnotes}
\usepackage{textcomp}
\usepackage{tabularx}
\usepackage{txfonts}
\usepackage{hyperref}
\usepackage{scrextend}
\addtokomafont{labelinglabel}{\sffamily\bfseries}
\usepackage{microtype}

\usepackage[multiple]{footmisc}

\usepackage{verbatim}
\usepackage{tikz}

\usepackage{placeins}

\usepackage[draft]{todonotes} 
\usepackage{authblk}

\newtheorem{myobs}{Observation}

\title{How not to establish the non-renormalizability of gravity}

\author[1]{Juliusz Doboszewski\thanks{jdoboszewski@gmail.com}}
\author[2]{Niels Linnemann\thanks{niels.linnemann@unige.ch}}
\affil[1]{Department of Philosophy, Jagiellonian University, ul. Grodzka 52, 31-044 Kraków, Poland}
\affil[2]{Department of Philosophy, University of Geneva, rue de Candolle 2, 1211 Genève 4, Switzerland}

\date{}

\begin{document}

\title{How not to establish the non-renormalizability of gravity\thanks{This work was partly performed under a collaborative agreement between the University of Illinois at Chicago and the University of Geneva and made possible by grant number 56314 from the John Templeton Foundation and its contents are solely the responsibility of the authors and do not necessarily represent the official views of the John Templeton Foundation. This work was also partly funded by a grant from the Swiss National Science Foundation (project number: $105212$\_$165702$).}
}



\maketitle

\begin{abstract} General relativity cannot be formulated as a perturbatively renormalizable quantum field theory. An argument relying on the validity of the Bekenstein-Hawking entropy formula aims at dismissing gravity as non-renormalizable \textit{per se}, against hopes (underlying programs such as Asymptotic Safety) that d-dimensional GR could turn out to have a non-perturbatively renormalizable d-dimensional quantum field theoretic formulation. In this note we discuss various forms of highly problematic semi-classical extrapolations assumed by both sides of the debate concerning what we call \textit{The Entropy Argument}, and show that a large class of dimensional reduction scenarios leads to the blow-up of Bekenstein-Hawking entropy.
\end{abstract}

%COMMENT_NL: perhaps we should remove d-dimensional as it is unimportant here?

\section{Introduction: could a theory of quantum gravity be yet another QFT?}

One of the central questions which have to be addressed at the beginning of the search for a theory of quantum gravity is the choice of theoretical framework. Could quantum gravity be simply yet another quantum field theory (QFT)? If not, which features of QFT need to go?

%COMMENT_NL: perhaps rather ,,A theory of quantum gravity" but I kin of agree with the sentiment that there should really be one (unless you did not intend to say this here of course)

Perturbative non-renormalizability of the Einstein-Hilbert action of general relativity is usually taken as a decisive argument against the viability of the standard QFT framework.\footnote{The perturbative non-renormalizability of gravity is typically only heuristically established in textbooks, that is by means of the power counting criterion of renormalizability (as in \citet{Zee}; see diagram 10.2 on p. 318 in \citet{PeskinSchroeder} for examples of how easily the power counting criterion fails). More accurately, one argues for gravity's perturbative non-renormalizability by pointing at the incurable infinities in loop diagrams of first order (in the matter case; second order in the matter-free case) \textit{while} expressing an expectation of infinitely more infinities at higher orders (cf. \citet{GoroffSagnotti}).}\footnote{Some authors, such as \citet{doughty1990lagrangian}, can even be interpreted as suggesting that there are two "unsatisfactory" features of classical general relativity: non-renormalizability and singularities. Whereas singularities have been thoroughly discussed in philosophy of science, non-renormalizability of GR has so far received at best passing attention.} Inadequacy of QFT, the standard story continues, forces us to choose among exciting and exotic options, including (among others) additional dimensions, fundamental discreteness, or holography.\footnote{There is much more to be said about the status of the alleged inadequacy of the framework, choice of features of QFT which are to be preserved (cf. for instance \citet{RovelliQFT}), and the importance of non-renormalizability for the theory choice and theory assessment; these philosophical considerations are crucial when it comes to the evaluation of exotic options we are seemingly faced with. But this is not our focus here.}

% Inclusion of GR as a dynamical component (as opposed to a fixed background) results in a theory with perturbative non-renormalizability, which in turn is understood as a breakdown of the framework. 

 %COMMENT_NL: ,,which are to be preserved". here we could give \citet{RovelliQFT} as an example. or do you think it is better to leave this undetermined? as you prefer.

But the perturbative notion of renormalizability is not the last word: there are multiple examples of theories which are perturbatively non-renormalizable but are renormalizable in a more general sense.\footnote{A standard example is due to \citet{Parisi}, who has shown that a perturbatively non-renormalizable four-fermion model is nevertheless describable up to high energies by a finite amount of parameters \textit{when} expanding it in the number of field copies rather than in a coupling constant as usually done.} In other words, even though the theory of quantum gravity cannot be a $4$-dimensional, perturbatively renormalizable quantum field theory, it could still be a $4$-dimensional, renormalizable quantum field theory -- generally construed. This question is particularly pressing in the light of research programs such as asymptotic safety, which aim at establishing non-perturbative renormalizability of GR through the hypothesis of a non-Gaussian fixed point in the renormalization group flow of the gravitational field. Evidence for existence of such fixed points is often hailed as a "good news" in the philosophy of physics literature (\citet{Butterfield}).

A natural question, then, is whether there could be a "no go" result, establishing non-renormalizability of GR in this more general sense. Primarily holographic aspects of gravity --- featuring in the black hole thermodynamic laws and gauge-gravity duality considerations --- are seen as incompatible with projects of turning GR into some kind of local quantum field theory. The Entropy argument --- to be dealt with in the following --- tries to turn this wide-spread sentiment into an \textit{explicit} argument against the renormalizability of $d$-dimensional General Relativity.

\section{The Entropy Argument}

The clearest exposition that we are aware of is given by \citet{Shomer}. Following Shomer, we stress that the core of this argument can already be found in earlier works, for instance in \citet{Banks}. It aims at establishing the claim that $d$-dimensional General Relativity cannot be turned into a renormalizable $d$-dimensional quantum field theory.

The argument relies on the incompatibility between the formulae of entropy of a generic conformal field theory (CFT) and that given by Bekenstein and Hawking (Bekenstein-Hawking entropy formula); accordingly, we will call it The Entropy Argument.

The argument runs as follows. Assume:

\begin{labeling}{Longer label\quad}

%COMMENT_NL: do you mean dimension-dependent or dimension-independent? I have changed it to dependent. also, it is not true that we only consider \Lambda=0, or? at least not in the following?

\item[(Bekenstein-Hawking)] The entropy of black hole states is given by the Bekenstein-Hawking entropy formula\footnote{In standard terminology, the Bekenstein-Hawking formula is relation between entropy and area, not energy.}; in particular\footnote{This formula only applies to spacetimes with a cosmological constant $\Lambda \leq 0$. Unfortunately, for the most interesting case of positive $\Lambda$ --- it is believed that we live in a universe with positive cosmological constant, after all -- there is no straightforward dimension-dependent formula available. Provided that talk about a de Sitter/CFT holographic scenario is physically sensible (for a critical take see for instance \citet{Strominger}), the Cardy-Verlinde entropy formula could be used (see \citet{Verlinde}).}

\[S\propto \begin{cases} E^{\frac{d-2}{d-3}} & \Lambda=0\\E^{\frac{d-2}{d-1}} & \Lambda<0\\ \end{cases},\]

where the spacetime dimension $d \geq 4$, $E$ denotes energy\footnote{Needless to say, black hole states are only expected to arise for sufficiently high spatial concentration of energy, as for instance upon \textit{probing} spacetime at sufficiently high energies with electromagnetic waves --- provided that the usual Planck-Einstein relation $E\propto G/R$ (where $R$ is the size of the region to be probed) holds.}, and $\Lambda$ the cosmological constant.

\item[(black hole dominance)] At high (probing) energies, the density of states is dominated by black hole states.\footnote{The concept of black hole dominance requires a notion of density of states, that is, it presupposes that different spacetime-matter-settings can be binned into the probing energy $E$ for which they occur.}

\item[(scaling)] At all energies, the same linkage between entropy, density of states, energy and other relevant dynamical parameters holds.

\item[(QFT-CFT link)] At high energies, every renormalizable $d$-dimensional QFT has approximately the same density of states as a $d$-dimensional CFT.

\item[(CFT entropy)] At high energies, the entropy of a CFT is given by $S \propto E^{\frac{d-1}{d}}$.\footnote{\label{footnote:derivation}This can be derived as follows. Assuming that (1) energy and entropy are extensive and that (2) temperature sets the dimensions in an otherwise scale-invariant theory, it follows that $S\propto R^{d-1} T^{d-1}$, and $E\propto R^{d-1} T^d$. Combining these two expressions, one finds that the entropy density $S/R^{d-1}$ scales with the energy density $(E/R^{d-1})^{\nu}$ where $\nu=\frac{d-1}{d}$. Precisely speaking, this gives a relation between entropy \textit{density} and energy \textit{density}. Taking $R$ to be independent of energy ($R(E)=R$), will render it as a relation between entropy and energy. This point will form an essential part of the criticism of the Entropy Argument in section \ref{CFTformula_density}.}
\end{labeling}

By \textit{black hole dominance}, the density of states at high probing energies is dominated by black holes states. By \textit{scaling}, the same remains true at the fundamental (Planck) scale. From \textit{Bekenstein-Hawking} and \textit{scaling} it follows that the density of states at the fundamental scale is given by the Bekenstein-Hawking formula. \textit{QFT-CFT link}, then, requires that the $d$-dimensional Bekenstein-Hawking entropy is approximately the same as the entropy of a $d$-dimensional CFT. From \textit{CFT entropy} a contradiction follows: the entropy within any renormalizable $d$-dimensional QFT will differ from the one given by the Bekenstein-Hawking entropy formula. So if a hypothetical gravitational QFT (for example, one obtained in the asymptotic safety approach) would reproduce a high energy behaviour in agreement with the Bekenstein-Hawking entropy formula, it cannot be renormalizable; and if it is renormalizable, then it cannot adequately reproduce the expected high energy behaviour in the light of the black hole dominance.

It is worth noting that Shomer and Banks do not take this to be an argument against the QFT framework \textit{per se}: the two formulae can agree if the gravitational QFT is in $d-1$ dimensions, which is exactly what happens in the holographic scenario in gauge/gravity duality. In this sense, then, the argument can be understood as establishing the holographic principle under the assumption of 'minimal' departure from the framework of QFT.\footnote{What counts as minimal departure from the QFT framework is of course far from being unambiguous -- simply observe how completely different frameworks for quantum gravity are each being advertised as being close to standard QFT.}

In what follows we do not want to take sides in the debates surrounding the fate of the Entropy Argument and programs such as asymptotic safety. Our aim is to trace out various ways in which problematic physical assumptions are used in the argument and reactions to it. After all, a logically correct argument is only as good as its premises. Before we discuss problems with individual premises in detail, we would like to highlight that the argument -- being built around the energy version of the Bekenstein-Hawking formula (as opposed to its more standard area formulation) -- commits one to a rather specific notion of entropy from the beginning.

\subsection{The notion of entropy}

The Entropy Argument builds on a version of the Bekenstein-Hawking formula in which the entropy of a black hole is expressed in terms of energy rather than area (see also footnote \ref{footnote:derivation}). From a statistical mechanical point of view one would consider this kind of entropy as linked to a system modeled as a microcanonical ensemble.\footnote{A generic microcanonical ensemble of particle configurations is characterized by a fixed total energy $E$, fixed particle number $N$ and a fixed volume $V$ with in which the particles move ($NVE$-ensemble). The entropy for such an ensemble is then defined as a function of the density of states which consequently is a function of $N$, $V$ and $E$. For a generic black hole, the microcanonical ensemble of states realizing the black hole is analogously characterized by a fixed energy, charge and angular momentum of the black hole. In particular, the entropy of a Schwarzschild black hole thereby becomes a function of energy.} Consequently, a black hole is hereby considered to be an isolated system with fixed energy which -- at least at the first sight -- runs counter to the strong expectation that black hole entropy is to a large extent due to entanglement effects between the interior and the exterior region of the black hole (for a review on entanglement entropy see for instance \citet{EntanglementReview}).\footnote{However, there may be ways to defuse this worry: for instance, \citet{RovelliVidotto} argue -- within the context of covariant loop quantum gravity -- that the microcanonical entropy on the one hand and entanglement entropy on the other hand can be understood as two sides of the same coin and are thus mutually compatible after all.}\footnote{Additionally, in certain situations the canonical or microcanonical ensemble description may fail to exist; an instructive example is discussed in \citet{Hawking}. Even though (at least at the thermodynamical limit) these two ensembles can and in practice are used interchangeably (see \citet{zwiebachchapter16}, \citet{wallaceblackholethermopart2}), one should, in general, make sure that in particular physical situations under consideration the necessary conditions for the existence of the canonical and/or microcanonical ensemble are satisfied.} Moreover, even though most of the times actual black hole-like objects are modeled as subregions of an asymptotically flat spacetime (which corresponds to being isolated from the environment), in realistic astrophysical situations spacetime metric is anything but not asymptotically flat. Justification for working with the microcanonical ensemble, then, seems to essentially rely on certain highly idealized representations.

In the set up of the argument above, it is moreover tacitly assumed that the entropy formula linked to a CFT and the Bekenstein-Hawking formula could and should be identified. One can see this as concealment of yet another premise, and one may thus question whether the two formulae in fact measure the same features of the system --- i.e. whether states counted by the two formulae are the same things (after all, what does it even mean for a CFT to have an entropy?). This is not a neutral premise: for example, one can object that the Bekenstein-Hawking entropy has nothing to do with the conventional understanding of entropy in statistical mechanics (but, say, is rather committed to an information-theoretic reading). 

Here, we would like to point out though that the so-called CFT entropy formula --- as shown by its derivation based on pure dimensional considerations before --- is supposed to provide the scaling behaviour of \textit{any} entropy (density) formula in terms of energy (density) and dimension set up in the context of that CFT. 

Still, it is worth asking whether entropy of certain types is sensible at high energies as a concept, which amounts once more to both a critical take on the validity of the Bekenstein-Hawking formula at high energies and a conceptual understanding of the interpretation of the formula.

\section{Reactions to The Entropy Argument}

Several lines of response to The Entropy Argument have been advanced. \citet{Basu2010} question the validity of the black hole-dominance conjecture (premise 2); several accounts (\citet{Laiho2011}, \citet{Coumbe2014}, \citet{Coumbe2015}) provide a dimensional reduction model in which the two formulae agree in the high-energies after all (a move we will interpret as being directed against scaling (premise 3)); \citet{Koch2013} and \citet{Falls} question the CFT entropy formula, albeit in different ways (premise 5). We will argue that almost all these lines share a problematic feature with proponents of the Entropy Argument, namely, a commitment to validity of a semi-classical approximation (a point we elaborate in \ref{conclusions}).

\subsection{Against the black hole dominance}

\citet{Basu2010} reduce premise 3, that is, the dominance of black hole states in high energy regimes (which they dub asymptotic darkness) to Thorne's hoop conjecture (cf. \citet{Thorne}), i.e. the conjecture that, given a region of characteristic size $R$ and energy $E$, a black hole will form if this characteristic size $R$ is smaller than the corresponding Schwarzschild-Radius $R_S$, i.e. $R<R_S=\frac{G}{E}$ where $G$ denotes the gravitational constant. In particular, any experiment of size $R$ and energy $E\propto \frac{1}{R}>E_{\text{Planck}}$ will lead to the formation of a black hole. The hoop conjecture builds in turn on a synthesis of a general relativistic reasoning (Schwarzschild Radius) and quantum mechanical reasoning (Planck-Einstein relation between size and energy in the experiment) which then gets extrapolated to the high-energy regime. In short,
\begin{enumerate}
\item
Asymptotic darkness builds on the hoop conjecture, and 
\item
the hoop conjecture holds as an extrapolation from the idea of black hole formation from low energy scales to high energy scales.
\end{enumerate}

As they point out, a derivation of asymptotic darkness would require "a particular theory of quantum gravity". So, asymptotic darkness is an extrapolation (and possibly and invalid one) of currently known physics to a range far beyond its typical domain of validity. The primary question is whether there are good reasons to accept 2., i.e. the extrapolation. This amounts to asking whether (1) the Planck-Einstein relation and (2) the Schwarzschild radius concept can be extrapolated to high-energies individually.

First, consider the Planck-Einstein relation. Provided that one accepts the necessity of a theory of quantum gravity (in the narrower sense of Quantum General Relativity or its curvature-corrected variant Quantum Gravidynamics\footnote{Which is an asymptotic safety term (see \citet{Niedermaier}) for any kind of curvature-corrected variant of GR where the standard Einstein-Hilbert action is extended by further terms depending on the curvature of spacetime, with the quantization of gravitational Hamiltonian which can include metric terms different from the Einstein-Hilbert action term and is the result of RG flow on the Einstein-Hilbert action from IR towards the UV.}), it seems unproblematic to assume the Planck-Einstein relation. This relation quantum-mechanically links concepts of size and energy with which one also expects to make contact in Quantum General Relativity or Quantum Gravidynamics. The use of the Planck-relation presupposes treating the metric field as a conventional physical field --- but this seems to be the case for Quantum GR/Gravidynamics in many ways anyway (for instance when considering the metric as subject to conventional quantization procedures or renormalization group flow). So, within a Quantum GR and its generalized variants, the Einstein-Planck relation might be extrapolated. However, we lack intuition about how to regard the status of this relation in the realm of approaches to quantum gravity in which a classical field theory of spacetime is seen as inappropriate as basis for quantization (this is not crucial at the moment, though).

Now, concerning the Schwarzschild radius: the occurrence of black holes (including corresponding singularities) is of course one of the major predictions of general relativity. It is now hoped that the nature of singularities linked to black holes gets explained by means of a theory of quantum gravity. But --- unlike for the Planck-Einstein relation (at least under the mentioned restrictions) --- we have no intuition at all why to trust the "Schwarzschild-radius formula" at very high energies. One of the merits of Basu and Mattingly's work is that they allow us for grounding this feeling by making explicit that it is genuinely open whether the relationship holds or not:

\begin{quote}
For a gedanken-experiment consisting of a ball of matter inside a volume $\Omega$ of proper size $L$ and energy $E \propto \frac{1}{L}$, the fixed point behavior of the running couplings conspire such that the necessity of forming a trapped surface inside $\Omega$ as we are shrinking the size of an experiment vanishes. 
\end{quote}

Note that the formation of a black hole requires the formation of a trapped surface. Thus, it is not necessary that high energy regions are dominated by black holes.\footnote{Basu and Mattingly conclude instead: "Hence the standard minimum length argument fails to necessarily be true in this case." This is not uncontroversial (at least from an operational point of view) since in a certain sense the asymptotic safety program always comes with a minimal length (\citet{Hossenfelder2013}, p.39):
\begin{quote}
It is essentially a tautology that an asymptotically-safe theory comes with this upper bound when measured in appropriate units.
\end{quote}
The discussion, then, seems to boil down to whether asymptotic safety can provide a workaround for the operational minimal length argument or not.}

Assuming symmetries\footnote{In case of stationary and perfectly symmetric matter, the hoop conjecture has been proven by \citet{Bizon}; but whether the proof generalizes if these symmetry assumptions are dropped is an open question.}, truncation and finitely many running couplings $C_{1}, ..., C_{n}$ with fixed points $c_{1}, ..., c_{n}$, one can find some numerical values of the fixed points $c_{1}, ..., c_{n}$ such that the required trapped surface forms, and some other numerical values of fixed points $c_{1}, ..., c_{n}$ such that trapped surfaces do not form. We know that the first set is non-empty (since in some cases the hoop conjecture has been proven), but we do not currently know whether the second set is non-empty. If the second set would be non-empty, then the black holes would not form, and black hole domination would not be true at arbitrary high-energies. Unless one obtains an estimate of the potential ranges of the values of fixed points which give rise to the formation of trapped surfaces, we should remain agnostic about Shomer's premise 3 even if the idealizations (stationary and perfectly symmetric matter) are granted, not to mention the more general case where these symmetries do not need to hold.

All of this takes us to the first observation we make:

\begin{myobs}
The Planck-Einstein relation and the hoop conjecture hold. (Semiclassical presupposition 1)
\end{myobs}

Note that in a sense this argument does not conclusively rebut the validity of premise 3. Again, one does not know whether the second set is really non-empty, and even if one shows that there is a range of numerical values for fixed points $c_{1}, ..., c_{n}$ such that trapped surfaces do not form, it is not clear that the numerical values of fixed points are physical. For example, if there were no trapped surfaces if and only if at least one fixed point has the value above $20$, and in all physically relevant cases all of the values of fixed points were below $5$, the fact that in general the black hole dominance fails would not be of much help against The Entropy Argument.

Moreover, a problematic classical assumption plays a role in this debate.

\begin{myobs}
There is a notion of energy under which the hoop conjecture holds, and that notion is applicable in a quantum gravitational context. (Classical presupposition)
\end{myobs}

The formulation of the hoop conjecture depends on notions such as "some energy $E$" or (in Thorne's original formulation) mass $m$ compressed into some region. But these notions are not unambiguous: there is no local energy in general relativity, and there is more than one proposal for a quasi-local notion (not to mention ADM or Bondi mass). In particular (see section 13.2.2 of \citet{szabados2009}) there are formulations of the hoop conjecture which can be proven using Brown-York energy but whose analogues for Kijowski–Liu–Yau or the Wang–Yau energies demonstrably do not hold. In that sense, it is already far from clear whether asymptotic darkness does hold in classical GR or not --- even before one starts extrapolating to high energy scales. 
 
\subsection{Against the scaling}

\label{againstscaling}

\citet{Laiho2011}, \citet{Coumbe2015} and \citet{Laiho2016} note that a dimensional reduction scenario in which the dimension of spacetime changes at high energies --- a phenomena routinely found in Causal Dynamical Triangulation (CDT) and Asymptotic Safety, cf. for instance \citet{Ambjorn} and \citet{Niedermaier}, respectively --- can provide an answer to The Entropy Argument. \textit{Prima facie} Shomer assumes that the spacetime dimension is constant, and an account is provided which does render change in dimension in such a way that it avoids contradiction between the two formulae. Dimensional reduction occurs for the so-called spectral dimension of spacetime, which is defined in terms of the return probability of a random walker (see for instance \citet{Coumbe2015}). Unlike the standard topological dimension, the spectral dimension can also take non-integer values. This objection to the Entropy Argument naturally hinges on the relevance of the spectral dimension for the scaling behaviour of the entropy formulae. So, The Entropy Argument assumes:

\begin{myobs}Dimension stays constant in the UV. (Semiclassical presupposition 2)\end{myobs}

For $\Lambda = 0$, \citet{Laiho2011} analyze Euclidean Dynamical Triangulation (EDT) with an additional measure term, and observe a reduction of spectral dimension to $d = 3/2$ in the UV. For that particular value of $d$, the Bekenstein-Hawking entropy formula coincides with that for a CFT. 
One can object to this on general grounds: the result is obtained for EDT, which is highly unphysical, because it works with an incorrect signature of the spacetime metric. Most problematically, EDT is not able to properly reproduce four dimensional relativistic spacetime at low probing energies (see for instance \citet{Coumbe2014}). 

\citet{Loll} directly analyze Causal Dynamical triangulation and find the spectral dimension to be $1.80 \pm 0.25$ for an allegedly canonical point in the de Sitter phase region within the two dimensional parameter space of CDT. \citet{Coumbe2015} question this result with a new study on three further points in the de Sitter phase area whose results rather hint at a reduced dimension of $3/2$ than $2$. % (for details, consider in particular table 2 in their paper).

Still, it is not straightforward to accept \citet{Coumbe2015}'s latest findings on dimensional reduction for CDT, even if their selected choice of parameters -- for which dimensional reduction to $3/2$ occurs -- might be more representative for the de Sitter phase region than the parameter set leading to a dimensional reduction to $1.80\pm0.25$ -- as considered by \citet{Loll}. First, there is the disagreement with competing accounts. Both asymptotic safety (see \citet{Niedermaier}) as well as toy model calculations on information gain from black holes in a dimensional reduction scenario (see \citet{Carlip}) suggest a dimensional reduction to $2$ rather than $3/2$. Secondly, there is the principal question of how far the spectral dimension should be considered to be of physical relevance (albeit \citet{Akkermans} result suggest that the spectral dimension can be seen as the relevant one for thermodynamic scaling in the case of photon gases).

\begin{figure}[htbp] 
  \centering
     \includegraphics[width=0.8\textwidth]{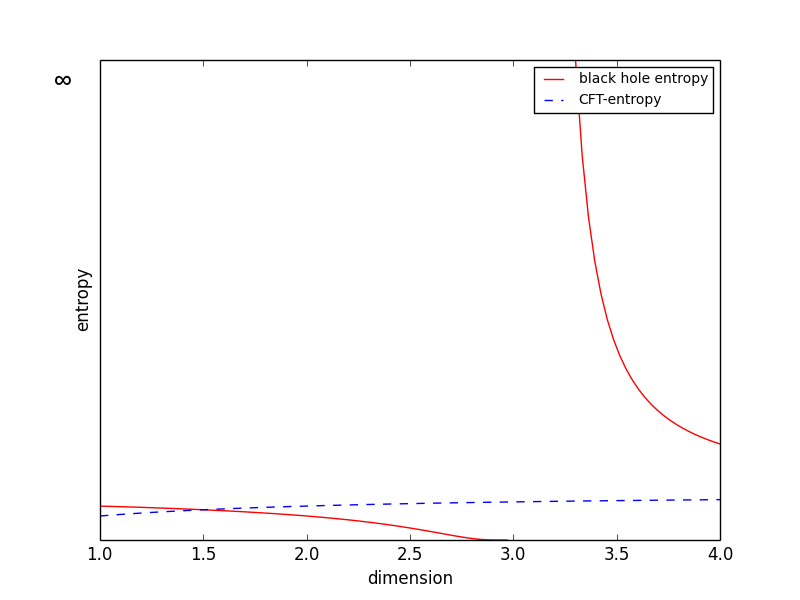}
  \caption{Schematic plot of CFT entropy and black hole entropy as functions of dimension, with $\Lambda=0$}
  \label{fig:Diagram1}
\end{figure}

A third, possibly more concrete worry about the usage of dimensional reduction in the context of the Entropy Argument comes from an observation about scale dependence of entropy (albeit limited to the $\Lambda=0$ case): $S_{CFT}$ and $S_{BH}$ agree at $d = 3/2$, but, as Shomer's analysis shows, disagree at $d = 4$. One can say that even if black hole dominance is taken for granted, disagreement between $S_{CFT}$ and $S_{BH}$ is irrelevant, since the discrepancy is visible only at low energies, and thus is negligible. Granted. Consider, however, $S_{CFT}$ and $S_{BH}$ as functions of $d$ in the range $[3/2, 4]$ (see figure \ref{fig:Diagram1}) --- now we take the dimensionality as a dynamical parameter of the theory. We see that:

\begin{myobs}Dimensional reduction is incompatible with the Bekenstein-Hawking formula.\end{myobs}

In more detail,
\begin{enumerate}
    \item $S_{BH}$ blows up at $d=3$, whereas $S_{CFT}$ is well-behaved throughout the whole range of $d$. The Bekenstein-Hawking entropy formula is thus unphysical as the dimension is being reduced to 3 or below (whatever the interpretation of the reduction would be), which seemingly puts doubt on the validity of the Bekenstein-Hawking formula in this approach.\footnote{Alternatively, one might say that black holes in lower dimensions have no well-defined entropy at all.}
        \item the dimensional reduction scenario is independent of the validity of the Bekenstein-Hawking formula in the sense that the scenario does not assume validity of the formula, but --- as long as dimensional reduction leads us to $d \leqslant 3$ --- it has to invalidate the formula!
    %\item the sum of $S_{CFT}$ and $S_{BH}$ is non-constant with respect to spacetime dimension $d$. Proponents of an information-theoretic reading of entropy should consider this as a problem; after all, in case of a scale-invariant theory, entropy (identified with information) should not change with resolution. 
\end{enumerate}

%AFTER_CQG_CORRECTIONS_NL: what information is encoded in Hawking radiation? isn't it thermal? how does it then allow you to probe a black hole???
% in how far is this point relevant anyway?

(The divergence of the Bekenstein-Hawking formula at $d=3$ (in the $\Lambda=0$ case) interestingly relates to the dimensional limitations on resolving a black hole as an outside observer (cf. \citet{Carlip}):
If dimensional reduction is real, then observers probing spacetime regions at different resolutions (i.e. different energy levels) will determine different (effective) spacetime dimensions for these regions. So, an observer studying a horizon region of a black hole of an evaporating black hole via its Hawking radiation will probe a smaller and smaller region of the ever decreasing black hole; and as probing ever smaller regions corresponds to probing at ever higher energies, the effective dimensions of these regions -- decreasing with energy -- will equally look smaller and smaller to her. 
For the $\Lambda=0$ case, Carlip and Grumiller find the Hawking evaporation of a (toy model) black hole\footnote{\citet{Carlip} use 2-dimensional dilation gravity to study an (Euclidean) Schwarzschild black hole in arbitrary dimensions.} to stop at a horizon scale corresponding to an effective dimension of $3$ in a dimensional reduction scenario. Thereby, observers outside the black hole will only be able to probe the black hole region up to a scale at which the (effective) spacetime dimension has got down to $3$ -- but not lower.)

%Both the divergence of the Bekenstein-Hawking formula and the halt of Hawking radiation at three dimensions hint at that a dimensional reduction scenario is most likely not compatible with a naive extrapolation of the Bekenstein-Hawking formula to higher dimensions.

%\citet{Carlip} that dimension should only go down to 2 and not to 3/2. This is based on a simple model
% this is built on a simple model (2-dimensional dilation gravity for (Euclidean) d-dimensional Schwarzschild black hole -- with and without cosmological constant. 
% it turns out that the smallest dimension up to which an observer can resovle will be two, but only (not 3/2) but only if she enters the black holes. otehrwise, depending on the cosmological cosntant value the obsrver who stays outside the balck hole will be only able to resolve dimsnions down between 2 and 3 for $\Lambda<0$, down to 3 for $\Lambda=0$ and greater than 3 for $\Lambda>0$ where Hawking evaporation stops. Arguably, \citet{Carlip}'s results are only of heuristic value.

All of this, of course, can be taken either against the validity of the Bekenstein-Hawking formula at high energies, or against the validity of the dimensional reduction view. Shedding more light on the meaning of the dimensional reduction would probably help: is it to be thought of as a physical process? Should $d$ be continuously parametrized? Or maybe only certain discrete values of $d$ should be considered? (That the Bekenstein-Hawking formula could be changed under RG flow will be featured below in section \ref{asymptoticallyimproved}.)

%AFTER_CQG_CORRECTIONS_NL: I have added a footnote on why fighting the Bekenstein-Hawking formula might amount to an uphil battle

\begin{myobs}Accepting dimensional reduction amounts to accepting Shomer's point.\end{myobs}

There is also a sense in which Laiho and Coumbe's position can be interpreted as admitting Shomer and Banks' point: $d$-dimensional QFT is insufficient, since a dimensional reduction mechanism is necessary to make the two formulae agree. Shomer notes that "to argue against the non-renormalizability of gravity is really to argue against the validity of the Bekenstein-Hawking formula, which is an uphill battle". The blow-up implied by dimensional reduction is naturally interpreted as an unphysical feature of the Bekenstein-Hawking formula; thus, it seems that any quantum-gravitational scheme in which dimensional reduction takes one from $d = 4$ to $d \leq 3$ has to fight such a battle.\footnote{Why this talk about an uphill battle? Now, the Bekenstein-Hawking entropy formula is not only a trusted result from black hole thermodynamics but it also provides the basis for one of the most dearly held principles -- at least within some communities -- towards a theory of quantum gravity, namely the holographic principle. Cf. \citet{Bousso}.}
Observe, however, that if the dimensional reduction scenario is true, the world is not described by a 4-dimensional local renormalizable QFT. In case of CDT, the theory is formally 4-dimensional (due to assumptions concerning types of the simplices used in the triangulation), but effectively is not 4-dimensional at high enough energies. In a clear sense, then, recourse to dimensional reduction amounts to admitting Shomer's point: at high energies the theory cannot be effectively 4-dimensional. Even worse, since dimensional reduction seems to come at a cost of violating Lorentz invariance (\citet{sotiriou2011spectral}), the world will not even be described by a standard local QFT at all!

Even if dimensional reduction is relevant to the scaling behaviour of entropy in terms of energy, it cannot help in dissolving the tension between the entropy formulae in the $\Lambda<0$ case. For no dimension $d$ will the scaling behaviours match, since the demand that $\frac{d-2}{d-1}=\frac{d-1}{d}$ --- as required for matching the two entropy formulae in the $\Lambda<0$ case --- cannot be fulfilled for any others numbers of $d$.

\subsection{Against the used entropy formula}

\label{againstCFT}

\subsubsection{Cardy-Verlinde formula}

The derivation of the CFT entropy formula used by Shomer builds on a CFT's scale-invariance but also on the assumption that entropy and energy are extensive functions of volume. Generally speaking, however, a CFT comprises a central charge which leads to a Casimir effect contribution to the total energy proportional to that charge and which thereby makes the entropy a non-extensive function of energy and volume (cf. \citet{Verlinde}). Within the current context of probing black hole structures, one could thus contest (as did \citet{Koch2013}) that the entropy formula put forward by Shomer —-- valid for a generic CFT in which the effect of the central charges is neglected —-- can still be used.\footnote{Another issue is that the derived CFT formula is a formula for the entropy \textit{density}, not the entropy itself. See section \ref{CFTformula_density}.}

Assuming that AdS/CFT correspondence (or some more general holographic scenario) holds, an alternative CFT formula is provided by the Cardy-Verlinde formula:

\[S=\frac{2 \pi}{n} R \sqrt{E_C (2E-E_C)}\] where $n$ is the dimension of the boundary space, $E$ the energy, $E_C$ the Casimir effect energy and $R$ the \textit{radius of the bulk structure}.

Even if the entropy formulae chosen by Shomer to account for the scaling of entropy in terms of energy are replaced by the Cardy-Verlinde formula, the respective CFT and black hole entropy formulae only match if the CFT is defined on the boundary of the bulk spacetime containing the black hole. Thus, accepting holography, it is still established that $d$ dimensional gravity cannot be turned into a $d$ renormalizable theory (while it might be subject to holographic renormalization). But this has the same problematic feature as the Entropy Argument itself: it may work for $\Lambda \leq 0$, but (in the absence of a worked-out de Sitter/CFT correspondence), how to understand it in the $\Lambda > 0$ case? Moreover, this makes the argument dependent on the truth of some holographic scenario. But what if this turns out not to be the case?

%[IT WOULD BE DESIRABLE IF THIS HELD: Still, the formula — once the energy dependence of R in a black hole scenario is taken into account — seems to have the same asymptotic scaling with energy as the standard CFT-formula.]

\subsubsection{Entropy density formula}

\label{CFTformula_density}
\citet{Falls} point out that Shomer’s CFT formula is strictly speaking relating energy density to entropy density (and not energy to entropy) (see the former, footnote \ref{footnote:derivation}). Consequently, the Bekenstein-Hawking entropy formula should be re-expressed as a formula of entropy \textit{density}, resulting in $s_{BH}\propto \epsilon^{1/2}$ for all dimensions $d$ where $s$ and $\epsilon$ denote entropy density and energy density respectively\footnote{Note that (1) $S \propto R^{d-2}$ and (2) $E \propto R^{d-3}$. From (2), we get (3) $R \propto E^{\frac{1}{d-3}}$. Now, $\frac{S}{R^{d-1}} \propto \frac{E^{\frac{d-2}{d-3}}}{R^{d-1}} \propto R^{d-1}$. As $\frac{E}{R^{d-1}} \propto R^{-2}$, the Bekenstein-Hawking entropy density $s_{BH}\propto \epsilon^{1/2}$ follows.}. The Entropy Argument suffers from:

\begin{myobs}Conflation of energy/entropy with energy/entropy density. (General naivety)\end{myobs}

\begin{figure} [h]
  \centering
     \includegraphics[width=0.8\textwidth]{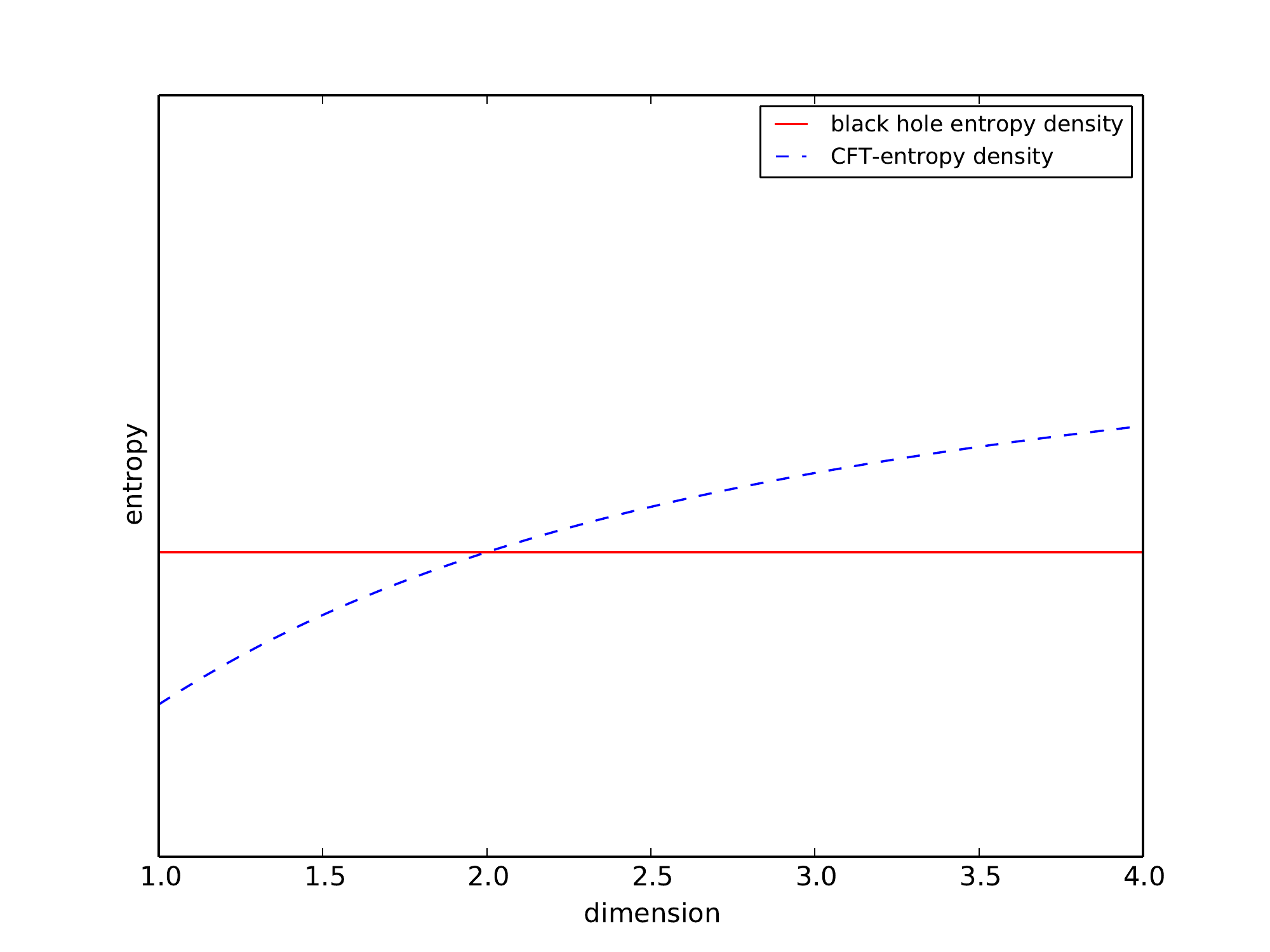}
  \caption{Schematic plot of CFT-entropy density and black hole entropy density as functions of dimension ($\Lambda=0$)}
  \label{fig:density}
\end{figure}

Then the problem of incompatible entropy density formulae can be solved for instance by (thermodynamically relevant) dimensional reduction to two dimensions (see Figure \ref{fig:density}) --- provided of course that the formula holds up to arbitrarily high energies. As pointed out before in section \ref{againstscaling}, several hints exist that this might be the case. And the Bekenstein-Hawking entropy density formula does not suffer from the blow-up problem of the Bekenstein-Hawking entropy formula at $d=3$ (see figure 2).

%It is thus an interesting interpretational question whether the Bekenstein-Hawking entropy density formula is somehow affected by the failure of the Bekenstein-Hawking formula of remaining finite at $d=3$.

\label{asymptoticallyimproved}

An asymptotic safety-improved Bekenstein-Hawking  entropy \textit{density} formula for the UV --- as calculated by \citet{Falls} ---  matches the CFT entropy density formula for $d=4$. The RG flow has the effect of letting the dimensionful Newtonian constant $G$ go to zero.\footnote{It is the dimensionless coupling constant $g=G k^2$ with respect to which asymptotic safety is said to have a non-trivial fixed point $g*:=\lim_{k \to \infty} g(k) \neq 0$ where $k$ is the energy scale.} As $G$ sets the scale, the absence of $G$ in the UV will render the theory at high energy scales as scale-independent so that the entropy density formulae for the black hole and for the CFT non-surprisingly match.\footnote{It would of course be interesting to know whether the match between entropy density formulae in the asymptotic safety improvement-scenario holds for $d$ dimensions (rather than just for $4$ dimensions) -- the correction from dimensional reduction would always work provided that the reduction is to $d=2$. Unfortunately we are not aware of any asymptotic safety-improved Bekenstein-Hawking entropy \textit{density} formula for higher dimensions.}

% in contrast to dimensional reduction idea even!

There seems to be no need to refer to dimensional reduction in this account. In fact, the account would not work if both the correction to the Bekenstein-Hawking entropy density formula and the correction to dimensional reduction were considered at once. How do these two approaches of letting the entropy density formulae match -- once via dimensional reduction, once via direct correction of the entropy density formula for the black hole --  relate to another? Are they two sides of the same coin, or rather mutually exclusive strategies?
It might seem that taking into account dimensional reduction while using the asymptotic safety-improvement of the Bekenstein-Hawking entropy density formula at the same time amounts to some kind of double correction for one and the same effect. 
But if the spectral dimension -- the dimension featuring in dimensional reduction -- really is the thermodynamically relevant one, then it seems even wrong not to account for the change in dimension in the entropy density formulae. The only way to accept \citet{Falls}'s resolution thus seems to dismiss the immediate relevance of the spectral dimensions for thermodynamic scaling.

\FloatBarrier

\section{Conclusions: on the semi-classical presuppositions}
\label{conclusions}

In the paper, we have assessed the following concrete replies to The Entropy Argument:

\begin{labeling}{Longer label\quad}
    \item[(Against the black hole dominance)]
    \citet{Basu2010} can be understood as trying to avoid the fallacy of assuming Black hole dominance \textit{while} still subscribing to the validity of the Bekenstein-Hawking formula at all energy scales at which black holes form. They reject operationalist implications of semi-classical approximations, i.e. approximations based on solutions to the semi-classical Einstein equations.
    
    \item[(Against the scaling)]
\citet{Laiho2011} and \citet{Coumbe2015} assume that they need to find a dimensional reduction scenario such that the reduced dimension $d$, after substitution, makes the Bekenstein-Hawking and CFT formulae equal. The Entropy Argument extrapolates from the currently known classical (or semi-classical) regime to a regime about which we can only surely know through a theory of quantum gravity, by assuming not only that the Bekenstein-Hawking formula is valid at intermediate scales (in which semi-classical theory is applicable), but also at truly quantum-gravitational scales. 
Considering dimensional reduction in the $\Lambda=0$ case, we discovered an interesting dichotomy: \textit{either} dimensional reduction holds at high energies while the Bekenstein-Hawking formula does not (The Entropy Argument cannot get off the ground anymore) \textit{or} the validity of the Bekenstein-Hawking formula at high energies rules out certain candidates of quantum gravity promoting dimensional reduction (as asymptotic safety or CDT). In the latter case, dimensional reduction cannot be used to defuse The Entropy Argument (interestingly, in that case, not only The Entropy Argument would still work against the validity of asymptotic safety itself, but would also show the impossibility of dimensional reduction --- a feature asymptotic safety and CDT are, after all, assumed to exhibit).
The straightforward dimensional reduction reply is not available in a $\Lambda<0$ scenario.

\item[(Wrong entropy formula)] \citet{Falls} rightly point out that the CFT entropy formula is in fact an entropy density formula. Taking into account asymptotic safety improvements to the Bekenstein Hawking entropy \textit{density} formula, it turns out that the entropy density formulae for the CFT and the black hole match. Alternatively, dimensional reduction to $2$ (as promoted by asymptotic safety) will lead to a matching of entropy density formulae. 
Their proposal of an RG-corrected entropy formula (which is incompatible with a scenario of thermodynamically relevant dimensional reduction and restricted to a $\Lambda=0$ scenario) would refute \citet{Shomer}'s claim that $d$ dimensional GR cannot give rise to a $d$ dimensional renormalizable QFT.

%But what could the justification of such a move be?

%COMMENT_NL: I have removed the Doplicher-like which was in front of fallacy    

%COMMENT_JD: ahh, there is one thing I am worried about. Is it OK to say "semi-classical approximation" in the context of Doplicher-like extrapolation? After all, I want to use "semi-classical" to mean "solution of semi-classical Einstein's equations".
%REPLY_COMMENT_NL: I do not think that this is such a big problem

%\citet{Koch2013} do not argue against extrapolations and semi-classical assumptions; the best interpretation of their position is that there is an unresolved in-house debate concerning the correct entropy formula for CFT.

%COMMENT_NL: please look at my alternative rendering of the two readings of Koch. I think I have a slightly different understanding of their argument. please tell me if you disagree with it, I will then immediately recreate the earlier version. otherwise, if you like it, I can adjust the section above accordingly
\end{labeling}

%COMMENT JD: now I think we should simply remove the following paragraph.
%In the end, it seems as if the Entropy Argument can be objected to in an unproblematic way by reference to asymptotic safety corrections to the scaling behaviour of quantities in the high energies. On the way to this, however, all sorts of semi-classical presuppositions about (1) the validity of the Bekenstein-Hawking formula, (2) the indifference between entropy and entropy density, but also (3) the putative constancy of spacetime dimensions came into view.

In a clear sense both the Entropy Argument and most replies to it are based on a sort of intuition pump. But this sort of intuitions seems dangerously unjustified when applied to energy scales far beyond the domains of the theories from which they do come from.

%NEW_AFTER_REVISION
% I check that O(N) nonlinear models, for d>2 are all renormalizable. see wiki for citations. YM-theories are not renormalizable from d>4 on, but are they all renormalizable in general?
% I have no idea how to back this up. perhaps we should leave this out after all?

%More, we have left out a possible criticism of the QFT-CFT link premise in the entropy argument: for both O(N) nonlinear sigma-models and Yang-Mill theories the QFT-CFT link is broken, i.e. at high energies the density of states of neither of these theories has approximately the same density of states as a CFT of the same dimension\footnote{Many thanks to Max Niedermaier for pointing this out to us.}

%\todo{PROPOSAL NL} 

%COMMENT_NL: now something on the black hole formula should follow again

% More generally, we should not only look at the validity of the premises on their own but how they are used within the inference of the argument.
%COMMENT_NL: actually, it seems to me that we really criticize tacitly assumed premises in addition to the ones given, rather than the inference in the argument (but yes, this probably amounts to the same thing)

The premise we find to be the most controversial in the Entropy Argument is the semi-classical presupposition of the validity of the Bekenstein-Hawking entropy formula \textbf{at high energy scales}, (or, in other words, exactness of validity of Bekenstein-Hawking entropy). This presupposition is present on both sides of the debate, but no justification is given for it: why exactly do we expect the Bekenstein-Hawking entropy to be a well-defined quantity at energy scales close to the Planck scale?
%JD: I am removing this part, because Bekenstein-Hawking entropy is a theorem in classical GR - Hawking radiation is semi-classical and what the analogue models are about! What is semi-classical, of course, is the extrapolation to high energies, Doplicher-style!

%At the same time, the Bekenstein-Hawking formula has in its lower energy variant --- if at all --- only been confirmed by means of analogue experiments, i.e. in the context of so-called analogue gravity (cf. \citet{Dardashti}) and it is actually the UV version we are interested in!

The Entropy Argument crucially depends on another assumption that a major feature of the high-energy behaviour of gravity can already be extrapolated from the classical theory, that is to say black hole dominance at high energies (premise 2). Moreover, it assumes scaling (premise 3), i.e. that the link between energy and density of states correctly captures an essential feature about the high energy regime of gravity---(again) something which in fact only a theory of quantum gravity can tell us about. These assumptions are equally extrapolated from classical theory, a fallacy similar to the one \citet{Wuthrich2005} discussed in the context of Doplicher's argument \citep{Doplicher} for the necessity of the quantization of the gravitational field.

\section*{Acknowledgements}
J. D.'s work was partly performed under a collaborative agreement between the University of Illinois at Chicago and the University of Geneva and made possible by grant number 56314 from the John Templeton Foundation and its contents are solely the responsibility of the authors and do not necessarily represent the official views of the John Templeton Foundation.

N. L. would like to thank the Swiss National Science Foundation for financial support (105212\_165702), and the Lorentz Center in Leiden for providing the platform for an enriching conference on Quantum Spacetime and the Renormalization Group. N. L. would also like to thank the the participants of this conference for valuable feedback.

Both authors would like to thank Claus Beisbart, Vincent Lam, Max Niedermaier, Carina Prunkl and Christian W\"uthrich.

\bibliographystyle{spbasic}      % basic style, author-year citations
\bibliographystyle{spmpsci}      % mathematics and physical sciences
\bibliography{ressources}   % name your BibTeX data base

\end{document}